# Tailoring the Energy Gap of Hybrid Hexagonal Boron Nitride Sheets Embedded with Carbon Domains of Different Shapes and Sizes

Cherno B. Kah, M. Yu and Chakram S. Jayanthi

Department of Physics and Astronomy, University of Louisville

**Abstract**

We present in this work the size-dependent and shape-dependent properties of carbon domains in hybrid *h*-BN/C sheets with the goal of tailoring their energy gaps. We have considered triangular, hexagonal, circular, and rectangular carbon domains embedded in *h*-BN sheets and optimized the structure of these *h*-BN/C sheets using the recently developed semi-empirical method, referred as SCED-LCAO. Calculated energy gaps of hybrid *h*-BN/C sheets exhibit interesting size- and shape-dependent properties. The energy gap of *h*-BN/C sheets embedded with rectangular domains shows a semi-metal behavior, while the size-dependence of the energy gap of hexagonal carbon domains exhibit a power-law decrease, and that of the energy gap of *h*-BN/C sheets with circular carbon domains show an oscillatory behavior. It is evident from the density of states calculations that the reduction in the energy gap of hybrid h-BN/C sheets compared to pristine *h*-BN sheets mainly arises from carbon domains. The bond distortions due to competing bond lengths (C-C, C-B, C-N, B-N, *etc.*) at the domain boundary also introduce states in the gap and this is particularly evident in sheets with smaller carbon domains.

I.     Introduction

With the advances made in synthesizing two-dimensional materials such as graphene and monolayer *h*-BN, there is unprecedented enthusiasm in the research on two-dimensional materials as they open the door to many technological innovations and provide a perfect playground to test the existing theories and models for two-dimensional systems. The properties of materials in reduced dimensions are different from their bulk counterparts because of effects arising from the quantum confinement of electrons and size-effects that profoundly alter the electronic, transport, optical, and other properties of 2D materials. Specifically, a fundamental research on 2D ternary systems is exciting because, by varying the percent composition of an element in a hybrid structure, one could tune the energy gap. Such materials design studies are necessary for the progress in photovoltaics research.

Graphene is known to have a unique band structure with a zero band gap at the Dirac point and a linear energy dispersion near its Fermi level [1], while *h*-BN is a wide band gap material with an energy gap of 5.9 *eV* [2, 3]. With synthesis of these materials via exfoliation or CVD becoming a routine, it is natural to ponder over the question: "whether it is possible to synthesize an atomic sheet composed of ternary elements B, N, and C?" If so, will it form an alloy sheet or will it phase separate into BN domains and graphene domains?

Recently, the Rice university group synthesized 2D hybrid structures of boron nitride and graphene domains using methane and ammonia-borane as precursors for BN and C in a thermal catalytic CVD method [4]. The experimental set-up allowed the carbon content in the hybrid structure to be controlled and thus allowing a study of the energy gap dependence as a function of the atomic percent composition of carbon (0%, 35%, 65%, 85%, etc.) in the hybrid structure. A variety of characterization tools ((AFM, HRTEM, Raman, XPS, EELS, and UV-visible absorption spectroscopy, *etc*.) was used to determine the atomic structure and bonding of hybrid ternary BNC films. Based on the results of these characterizations, the authors argue that the observed structure was neither an alloy nor a stacked structure of *h*-BN and graphene but a hybrid 2D structure composed of graphene and BN domains. Specifically, the absorption spectroscopy experiment conducted on the synthesized samples reveal a second absorption peak corresponding to carbon domains in the BNC sheet and, furthermore, the position of this peak shifted to higher wavelengths as the carbon content in the hybrid structure increased. More specifically, the optical gap shifted from 1.62 eV to 1.51 eV, as the carbon content in the hybrid BNC sheet increased from an atomic percentage of 65% to 85%. This experiment demonstrates that by controlling the domain size of graphene domains, one can tune the energy gap in a hybrid BNC sheet composed of BN and graphene domains. Other experimental groups also reported similar results [5-7]

On the theoretical front, the literature reports a study that calculates the formation energies and energy gaps of circular shaped carbon quantum dots (QD) of different diameters embedded into *h*-BN sheets using the DFT method as implemented in the SIESTA code [8]. This study finds the formation energy to oscillate as a function of the size of the carbon quantum dots and concludes that the hybrid system has lower energy whenever the quantum dot contains complete aromatic rings. The density of states for the hybrid (QD + *h*-BN) system seems to indicate the existence of a strong hybridization among 2p orbitals of B, N and C. The energy gap dependence as a function of the diameter of graphene QD, as obtained in Ref. [8]. Similar results were also reported by other theoretical groups [9, 10]

It is worth noting that the theoretical study presented in ref. [8] is restricted to circular graphene quantum dots in *h*-BN sheets and the highest atomic percentage considered is less than 11%, while the BNC films investigated by the Rice group contain graphene domains and BN domains of random size and arbitrary shapes.

Motivated by the above two studies, we have performed a systematic study of different shaped (triangular, rectangular, circular, hexagonal) and sized graphene domains (up to an atomic percentage of 40%) in *h*-BN sheets to understand the energy gap-dependence of the ternary BNC sheet as a function of the size and shape of graphene domains. In section 2, we will briefly introduce our calculation method and in section 3, we will discuss the results obtained from such a study. Finally, we will give our conclusion in section 4

II. Calculation method

The computations are performed using a semi-empirical Hamiltonian, referred as SCED-LCAO, developed at the University of Louisville [11]. In this approach, the screening effects of electron-electron and electron-ion interactions in a multi-atom environment are captured through environment-dependent (ED) functions and charges are calculated self-consistently (SC). The framework of our approach uses a linear combination of atomic orbitals (LCAO) and a minimal basis set (sp$^3$) to describe the Hamiltonian. While the SCED-LCAO Hamiltonian mimics the interaction terms of the density functional theory (DFT) Hamiltonian, it is parameterized. The matrix element of the new generation of SCED-LCAO the Hamiltonian is given as [12]:

$$\hat{H}_{i\alpha,j\beta} = \frac{1}{2}[\varepsilon'_{i\alpha} + \varepsilon'_{j\beta}]K(R_{ij})S_{i\alpha,j\beta}(R_{ij}) + \frac{1}{2}[\sum_{k\neq i} W_{i\alpha}(R_{ik}) + \sum_{k\neq j} W_{j\beta}(R_{jk})]K(R_{ij})S_{i\alpha,j\beta}(R_{ij}) + \frac{1}{2}[(N_i - Z_i)U_i + (N_j - Z_j)U_j]S_{i\alpha,j\beta}(R_{ij}) + \frac{1}{2}[\sum_{k\neq i}\{N_k V_N(R_{ik}) - Z_k V_Z(R_{ik})\} + \sum_{k\neq j}\{N_k V_N(R_{jk}) - Z_k V_Z(R_{jk})\}]S_{i\alpha,j\beta}(R_{ij})$$  (1)

The Hückel energies terms $\varepsilon'_{i\alpha}$ and $\varepsilon'_{j\beta}$ in Eq. 1 are adjustable parameters. The first term (in between square brackets) in the expression of the matrix elements of the Hamiltonian includes the two-center interaction term followed by the expression with the $W$ term that takes account the possible occupation of the excited local orbitals in an atomic aggregation and therefore, the effects of interactions with neighboring atoms. The on-site electron-electron interactions (third term in between square brackets) provide the framework for the charge redistribution calculation and the fourth terms take care of the environment dependence and multicenter interactions also called the environment dependent term defined by the different potential functions $V_N$ and $V_Z$. The optimized SCED-LCAO Hamiltonian parameters for nitrogen are listed in Table 1. Calculated cohesive energies and structure properties of various $B_nN_m$ and $C_nN_m$ clusters are consistent with the DFT results [13]. Robust test on h-BN sheet and Wurtzite BN (w-BN) bulk structure also show its transferability, namely, the optimized lattice constant is about 3.5% (4%) overestimated for h-BN (w-BN) compared to the experimental values [3, 14], and calculated energy gap is 3.303 eV (3.701 eV) for h-BN (w-BN), which underestimated compared to experimental measurements (5.9$eV$ [15, 16] and 5.4$eV$ [2], respectively for h-BN, and 5.4 eV for w-BN [3]).

Table 1 SCED-LCAO Hamiltonian parameters for Nitrogen element.

| Hamiltonian Parameters for Nitrogen | | Hamiltonian Parameters for Nitrogen | |
|---|---|---|---|
| $\varepsilon_s(eV)$ | -26.233 | $U(eV)$ | 14.378 |
| $\varepsilon_p(eV)$ | -13.842 | $B_Z(Å^{-1})$ | 3.546 |
| $\varepsilon'_s(eV)$ | -29.806 | $A_N(eV)$ | -2.793 |
| $\varepsilon'_p(eV)$ | -21.139 | $B_N(Å^{-1})$ | 2.014 |
| $W_s^0(Å^{-1})$ | 0.588 | $\alpha_N(Å^{-1})$ | 3.881 |
| $W_p^0(Å^{-1})$ | -0.355 | $d_N(Å)$ | 1.035 |
| $\alpha_{s,w}(Å^{-1})$ | 1.481 | $\alpha_{NC}(Å^{-1})$ | 0.439 |
| $\alpha_{p,w}(Å^{-1})$ | 1.805 | $\alpha_{NB}(Å^{-1})$ | 0.591 |
| $\alpha_K(Å^{-1})$ | 0.359 | | |
| Overlap Parameters for Nitrogen | | Overlap Parameters for Nitrogen | |
| $B_{ss\sigma}(Å^{-1})$ | 0.602 | $B_{pp\sigma}(Å^{-1})$ | -1.518 |
| $\alpha_{ss\sigma}(Å^{-1})$ | 1.778 | $\alpha_{pp\sigma}(Å^{-1})$ | 3.412 |
| $d_{ss\sigma}(Å)$ | 0.262 | $d_{pp\sigma}(Å)$ | 0.948 |
| $B_{sp\sigma}(Å^{-1})$ | 0.460 | $B_{pp\pi}(Å^{-1})$ | 0.032 |
| $\alpha_{sp\sigma}(Å^{-1})$ | 3.017 | $\alpha_{pp\pi}(Å^{-1})$ | 3.769 |
| $d_{sp\sigma}(Å)$ | 1.221 | $d_{pp\pi}(Å)$ | 0.707 |

Figure 1 depicts the four different types of graphene domains embedded in an *h*-BN sheet. In addition, the size of the graphene domains will be changed to understand the roles played by both size and shape of graphene domains in altering the band gap of pristine *h*-BN sheet. As a first step, we optimize the structure of a pristine *h*-BN sheet and determine the energy band gap as obtained by the SCED-LCAO method. The Hamiltonian parameters for nitrogen as obtained in chapter IV are used in constructing the SCED-LCAO Hamiltonian for the *h*-BN system. A supercell of size 15x10 containing 600 atoms and a vacuum of 1000 Å is used in our SCED-LCAO calculations. Simulations of pristine *h*-BN sheets were followed by simulations of 2D hybrid BNC system, containing graphene domains of different shapes (triangular, hexagonal, circular, and rectangular) and sizes embedded into the *h*-BN sheet. In the case of triangular graphene domains embedded into *h*-BN sheets, we distinguish two cases based on how the etching is performed. In the first case, the etching of *h*-BN sheets is such that we replaced all boron atoms falling along the perimeter of the triangular etch by carbon atoms (T-B type graphene domains) and in the second case, we replaced all nitrogen atoms falling along the perimeter of triangular etch by carbon atoms (T-N type graphene domains). In the former case, the interface between the graphene domain and the *h*-BN sheet will have C-N bonds while in the second case, the interface between the graphene domain and the *h*-BN sheet will have C-B bonds. It will be interesting to investigate how this difference in the bonding characteristics of the interface will influence the electronic density of states and the energy gaps.

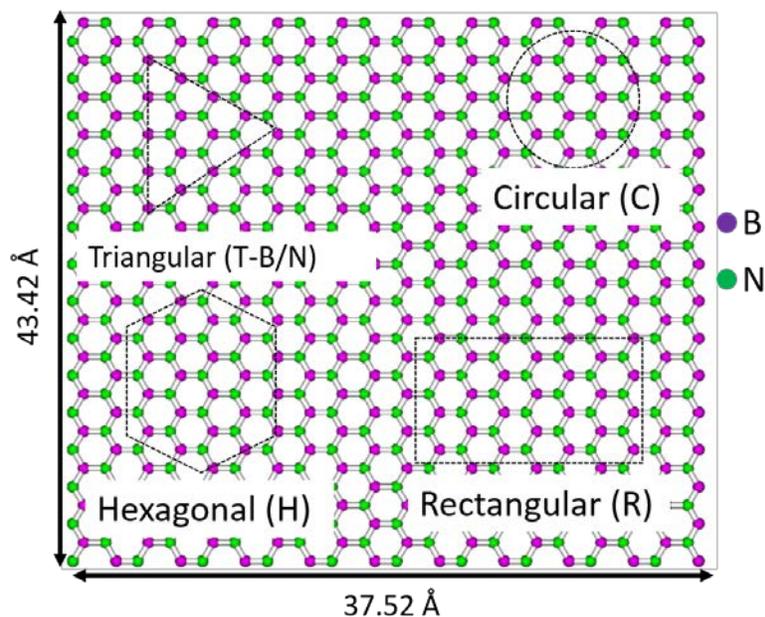

Figure 1 A schematic of a 15x10 *h*-BN sheet and the four types of graphene domains (triangular, circular, hexagonal and rectangular) embedded in pristine *h*-BN sheet.

III. Results and Discussion

A. Triangular shaped graphene domains in hybrid h-BN/C sheets.

A-1. Graphene domains with C-N interface (T-B).

Triangular shaped graphene domains are embedded in the 15x10 *h*-BN sheets. A total of 4 different sizes for the triangular domains were considered with an increase number of Carbon atoms as the sizes increase. Triangular 1 (T-$B_1$) has a total of 3 Boron atoms replaced by Carbon and it is the smallest in size with the 3 atoms forming the triangular shape. The next size is triangular 2 (T-$B_2$), a total of 12 Boron is replaced along the boundaries of the triangle to form the 3 different sides with 5 atoms per side. The total number of carbon atoms is 21. Triangular 3 (T-$B_3$) accounts for 21 Boron replaced along the perimeter of the domain with the total 57 carbon atoms in the domain, and triangular 4 (T-$B_4$) has a total of 30 Carbon atoms replacing 30 Boron atoms along the perimeter with the total 111 carbon atoms in the domain. These four samples are fully optimized and relaxed using the SCED-LCAO method and only two of them (i.e., T-$B_1$ and T-$B_2$) are found stable. They were shown in Fig. 2 (a).

*h*-BN has a wide band gap with the Fermi level clearly shown between the top of the valence band (VB) and the bottom of the conduction band (CB) (see the top panel in Fig. 2 (b)). Embedding a triangular graphene domain on the *h*-BN sheet changes the physiognomy of the density of states. For instance, in the cases of triangular Boron 1 and 2 (T-$B_1$ and T-$B_2$), even though the shapes and the positions of the VB and CB almost unchanged, there are a few 'impurity' states appearing within the gap of the pristine *h*-BN sheet and the larger the size of the graphene domain, the more the states, and therefore the narrower the band gap (see the middle and bottom panels in Fig. 2 (b)). With the appearances of some states between VB and CB of the DOS calculations, we were interested in finding the origin of such states hence it was proposed to calculate the local density of states (LDOS) located at the carbon domain. The bottom panel in Fig. 2 (b) shows the LDOS (red curve) for the biggest triangular domain (T-$B_2$) together with the total DOS for T-$B_2$ (blue curve). By comparing the total DOS (blue) with the LDOS (red), we found that the states appearing between the VB and CB are indeed contributions from the graphene domain.

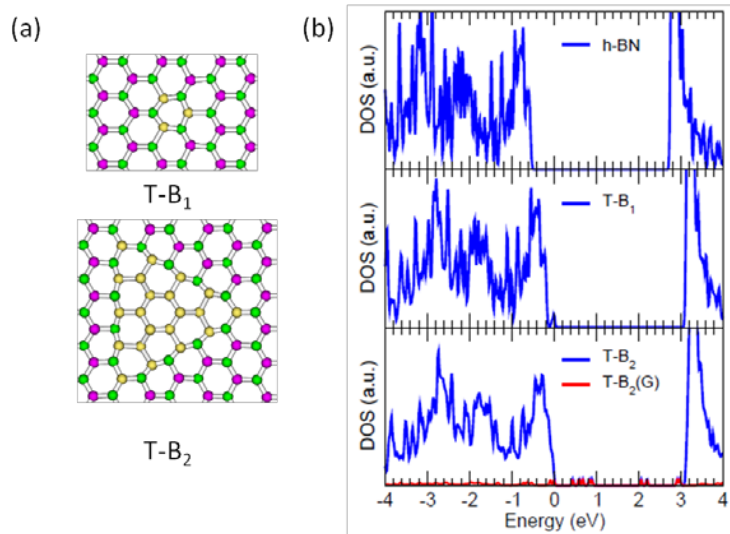

Figure 2 (a) The relaxed structures for *h*-BN/C hybrid sheets for two different sizes of triangular carbon domains, labeled T-$B_1$ and T-$B_2$, respectively with their initial structures obtained by replacing either 3 or 21 boron atoms within and along the perimeter of a triangular region of a pristine *h*-BN sheet by corresponding number of carbon atoms. The three panels in (b) correspond to the total density of states for a pristine *h*-BN sheet and the hybrid *h*-BN/C sheets labeled T-$B_1$ and T-$B_2$, respectively. T-$B_2$ (G) (red curve) represents the local density states from only the carbon domains. To make the comparison of different cases easier, we have plotted the density of states as a function of $\varepsilon - \varepsilon_F$ so that the Fermi level is always located at zero.

A-2. Graphene domains with C-B interface (T-N).

Similar as in triangular Boron, triangular Nitrogen also has four different sizes of triangular graphene domains. Triangular Nitrogen1 (T-$N_1$) were constructed with 3 nitrogen atoms replaced by 3 carbon atoms along the perimeter of the triangle which also is the smallest possible triangular domain; triangular nitrogen 2 (T-$N_2$) with 12 nitrogen atoms being replaced by carbon; triangular nitrogen 3 (T-$N_3$) in which 21 carbon atoms replaced 21 nitrogen atoms along the perimeter and finally triangular nitrogen 4 (T-$N_4$) where a total of 30 nitrogen atoms are replaced by carbon. These samples were optimized and fully relaxed using the SCED-LCAO, and the relaxed *h*-BN sheet with different size of triangular nitrogen graphene domains were shown Fig. 3 (a). The electronic density of states calculations for triangular Nitrogen show that the replacement of boron and nitrogen atoms by carbon atoms through embedding leads to the appearing of states both at the top of the valence band and the bottom of the conduction band (see Fig. 3 (b)) hence a decrease in the energy difference between the two bands. With increasing the size of the graphene domain, the shapes of VB and CB change, e.g., their peaks were broaden and their tails were extended towards the gap, leading to narrowing the band gap. The calculation of the LDOS located at the carbon domain for T-N$_4$ as an example is shown in Fig. 3 (b) and it reveals that the states appearing between the VB and CB are partially due to the presence of the carbon domain and partially due to the boron and nitrogen atoms at the boundary. The states are more pronounced as the size of the domain increases.

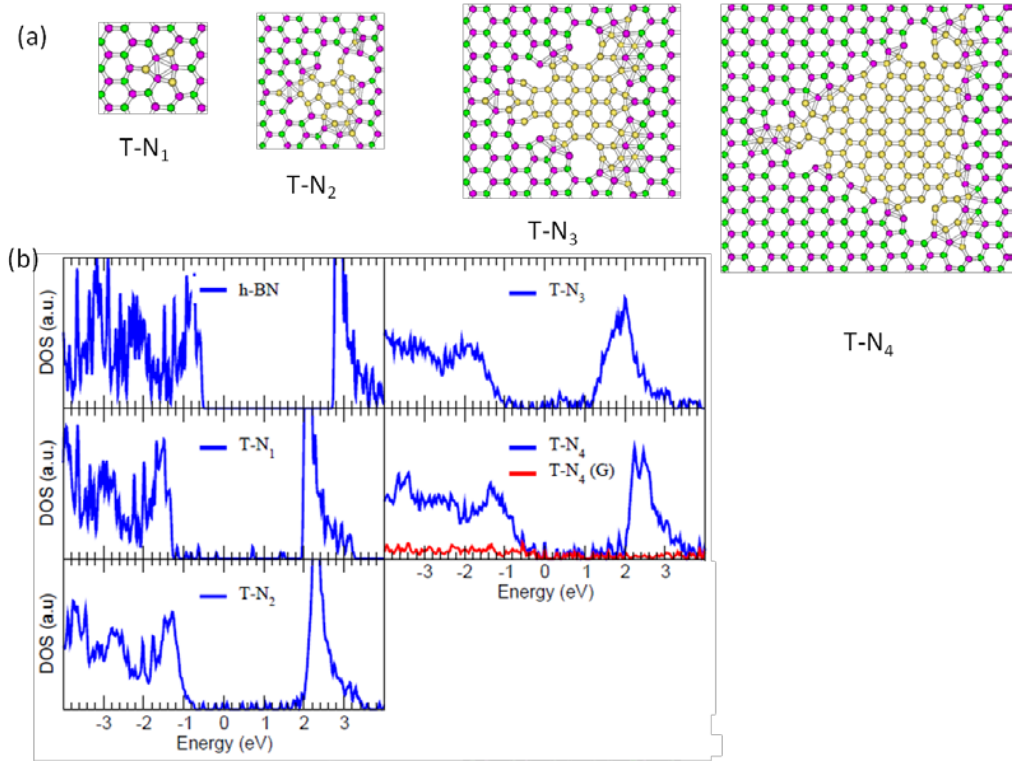

Figure 3 (a) The relaxed structures for four different sizes of carbon domains in *h*-BN/C hybrid sheets (T-N$_1$, T-N$_2$, T-N$_3$ and T-N$_4$) with their initial structures obtained by replacing (3, 12, 21, or 30) nitrogen atoms within and along the perimeter of a triangular region of a pristine *h*-BN sheet by corresponding number of carbon atoms. The five panels in (b) correspond to the total density of states for a pristine *h*-BN sheet and the hybrid *h*-BN/C sheets labeled T-N$_1$, T-N$_2$, T-N$_3$, and T-N$_4$, respectively. T-N$_4$ (G) (red curve) represents the local density states from only the carbon domains.

    B. Circular Shapes of graphene domains.

Circular graphene domains are constructed by creating the graphene domain with circular shapes embedded in our *h*-BN sheet. The possible smallest domain is circular1 (C-1) which is a complete hexagon of 3 boron and 3 nitrogen atoms which are all replaced by carbon hence a total of 6 carbon atoms. A total of 6 different sizes of circular shape (i.e., C-1, C-2, C-3, C-4, C-5, and C-6) were found to be possible to be embedded on our 15x10 BN sheet. The largest circle has a diameter of 23.97Å. It was found that, during the relaxation, the graphene domain kept intact towards the center of the circle, while at the boundaries the strong interactions between the 3 different atoms lead to break of the hexagonal symmetries locally more noticeable in the smaller circular domains. The relaxed *h*-BN sheet with various circular graphene domains were shown in Fig. 4 (a). The electronic density of states shown in Fig. 4 (b), predict the appearance of new states both at the top of the VB and also bottom of the CB similar to the cases found in the triangular domains. Through local density of states calculations (e.g., the red curve of Fig. 4 (b) shows the LDOS for graphene domain C-6), the states were identified to be mainly contribution from the graphene domain which is also observable from the total DOS since as the diameter of the circles increase the states also appear more clearly together with a shift of the Fermi level towards to the conduction band in the *h*-BN sheet; suggesting a decrease in the energy gap.

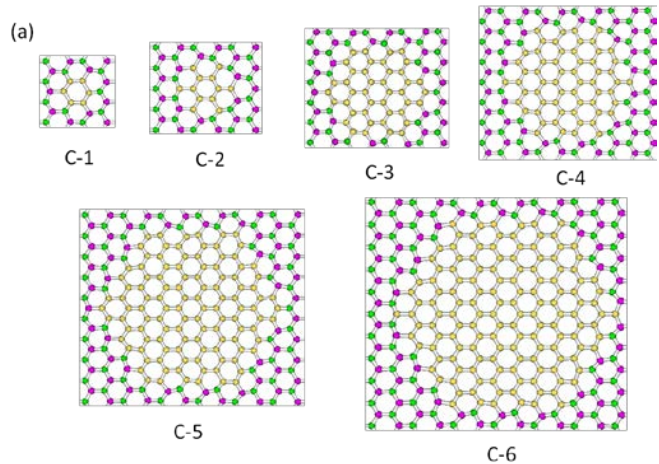

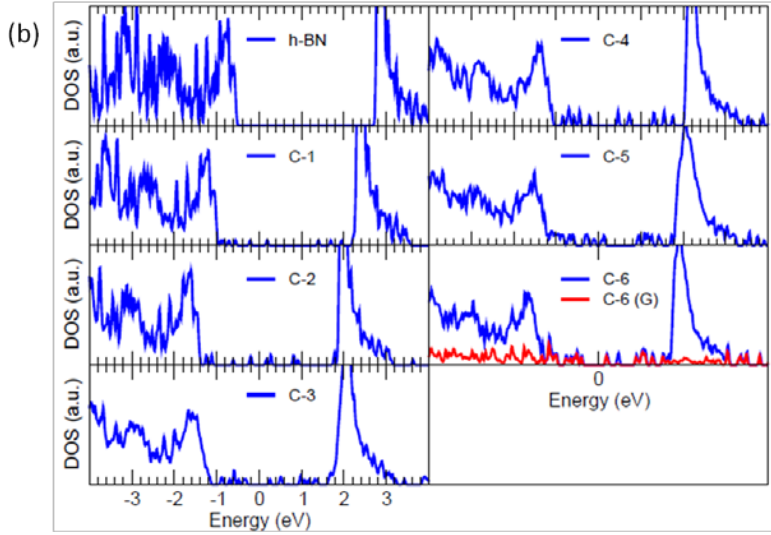

Figure 4 (a) Relaxed structures of *h*-BN/C hybrid sheets corresponding to six different circular graphene domains of different sizes. The seven panels in (b) correspond to the total density of states for a pristine h-BN sheet and the hybrid h-BN/C sheets labeled C-1, C-2, C-3, C-4, C-5, and C-6, respectively. C-6 (G) (red curve) represents the local density states from only the carbon domains.

C. Hexagonal Shapes of graphene domains.

Four hexagonal domains were considered which were constructed by the number of hexagons and are named as H-1, H-2, H-3, and H-4, respectively. The smallest hexagonal domain H-1 was formed by three nearest hexagons with thirteen carbon atoms. The second one, H-2 was constructed by seven nearest hexagons with six of them around the central one. When adding another twelve hexagons surrounded to the H-2, we formed H-3. The biggest hexagonal Carbon domain, H-4, was constructed by adding 18 hexagons at the boundary of H-3, with possible distance between two edges of the hexagon being 17.99Å. We found that, at the center of a given domain the graphene structure is well-defined. But as one moves from the center towards the hexagonal boundaries, the interaction between the carbon atoms and boron or nitrogen atoms becomes stronger with different bonds between two types of carbon. The two types of carbon atoms identified at the boundaries are $C_1$ and $C_2$. In $C_1$ the carbon is connected to two carbon atoms and a single nitrogen atom while in $C_2$ the carbon atom is connected to two carbon atoms and a single boron atom. These two types of carbon atoms are alternately distributed along the 6 edges of the hexagon domain. The relaxed *h*-BN sheet with different size of hexagonal graphene domains were shown in Fig. 5 (a). As in the previous discussions, the electronic density of states revealed the appearance of some hybrid states at the top of VB and the bottom of CB, leading to a decrease in their energy difference but the shape of the valence/conduction band almost remains the same as the pure *h*-BN.

The calculation of the LDOS for the hexagonal graphene domain also revealed that the states are originating from the carbon domains of the different hexagonal shapes. Fig. 5 (b) as an example, shows the total DOS of H-4 *h*-BNC (blue curve) with the LDOS for H-4 domain (red curve).

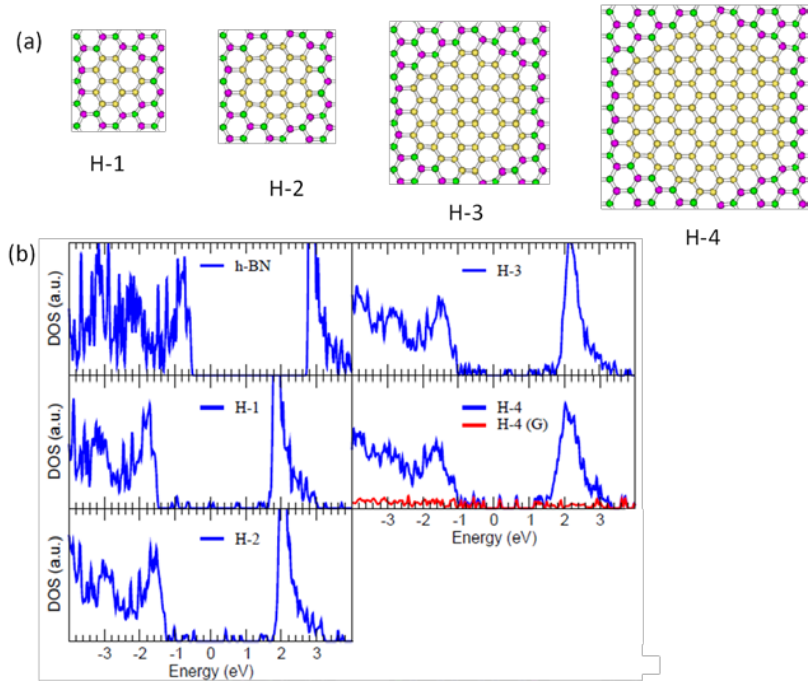

Figure 5 (a) Relaxed *h*-BN/C hybrid sheet corresponding to hexagonal graphene domains of different sizes. The five panels in (b) correspond to the total density of states for a pristine h-BN sheet and the hybrid h-BN/C sheets labeled H-1, H-2, H-3, and H-4, respectively. H-4 (G) (red curve) represents the local density states from only the carbon domains.

D. Rectangular Shapes of graphene domains.

Rectangular shapes of graphene domains are constructed and embedded in the *h*-BN sheet to create the *h*-BNC sheets and subsequently check the effect on the gap energy. A total of 7 rectangular shaped graphene domains were created, named as R-1, R-2, R-3 R-4, R-5, R-6, and R-7. Note that CB and CN bonds are located at opposite sides of the rectangular and alternatively distributed at the other two sides. The smallest rectangular graphene domain is rectangular 1 (R-1), in which a total of 6 carbon atoms substitute 3 boron and 3 nitrogen atoms and form the perimeter. The largest possible rectangular carbon domain for which there is no interaction between the neighboring domains is rectangular 7 (R-7), where a total of 42 carbon atoms substitute 21 boron and 21 nitrogen atoms along the edges of the perimeter with lengths of 29.97Å and 21.00Å, respectively. At the boundaries the interactions between the carbon atoms and boron or nitrogen atoms are strong but the symmetries are kept. The relaxed *h*-BN sheet with various rectangular graphene domains were shown in Fig. 6 (a). The electronic density of states calculations also revealed that the introduction of rectangular graphene domains within the *h*-BN sheet favored the appearance of some hybrid states at the top of the VB and the bottom of CB. As the area of the rectangular domains increases a decrease in the energy difference between the two bands was found when compared to the energy gap of pure *h*-BN sheet. The origin of the hybrid states between top of the VB and bottom of the CB were also confirmed to be from the carbon domains through calculations of the LDOS for the rectangular graphene domain (e.g., the red curve in Fig. 6 (b) shows the LDOS for R-7 domain).

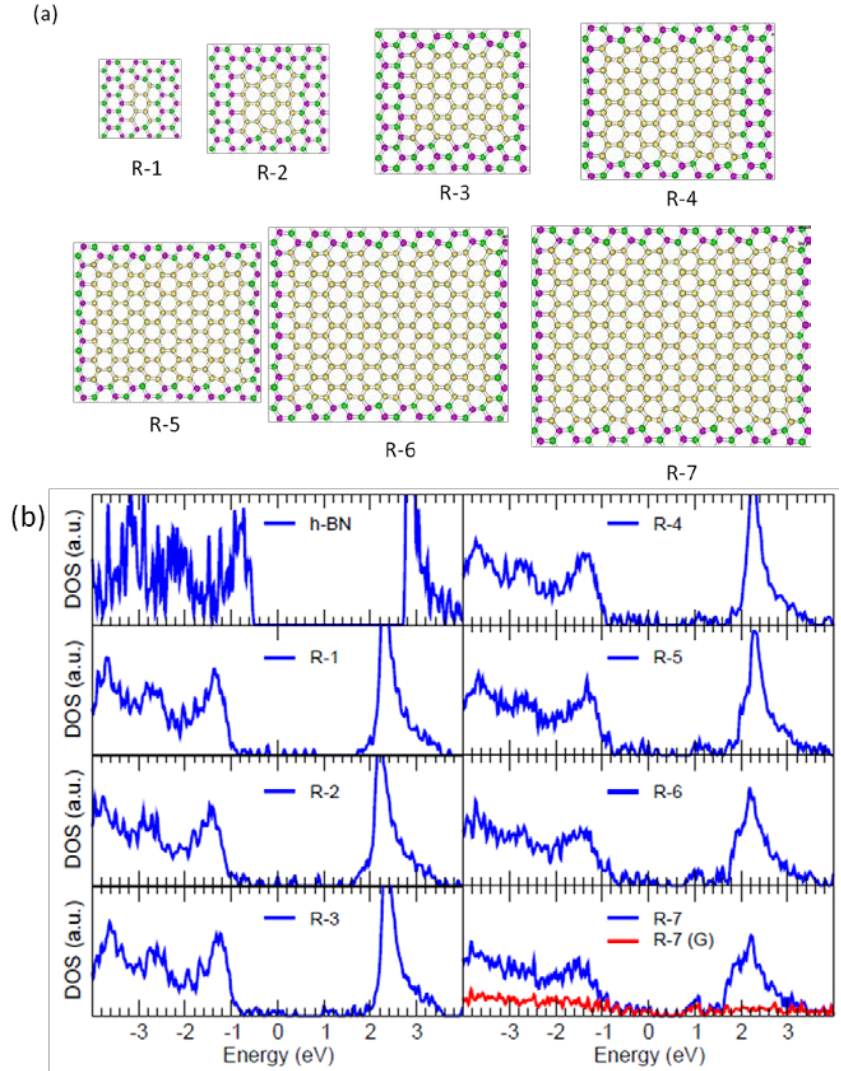

Figure 6 (a) Relaxed structures corresponding to rectangular *h*-BN/C hybrid sheets for rectangular graphene domains of different sizes. The eight panels in (b) correspond to the total density of states for a pristine h-BN sheet and the hybrid h-BN/C sheets labeled R-1, R-2, R-3, R-4, R-5, R-6, and R-7, respectively. R-7 (G) (red curve) represents the local density states from only the carbon domains.

E. Energy gap

Fig. 7 shows the calculated energy gap as a function of the area of graphene domains corresponding to different shapes in hybrid *h*-BN/C sheets. The values of the energy gap for all *h*-BN sheets with different shapes and sizes of graphene domains are listed in Table 2. For the case of Triangular-B graphene domains (black dots) with boron atoms at the boundary replaced by carbon, there exits only C-N bonds (~ 1.2 Å) at the boundary. On the other hand, since the bonds that exist within the boundary are C-C bonds (~1.46 Å), and those existing outside the boundary are B-N bonds (1.5 Å), a mismatch among the three types of bond lengths in the vicinity of the boundary causes a local strain in the system. This strain increases as the domain size increases, and we find that only the two smallest Triangular-B domains (i.e., T-$B_1$ and T-$B_2$) are stable in our SCED-LCAO simulations. There is an empty defect state induced by carbon domain which is about 0.048 eV above the top of VB in T-$B_1$. Except this defect state, the gap between VB and BC is about 3.0 eV, comparable to the pristine *h*-BN sheet (3.3 eV), since the substitution of boron by carbon atoms takes place at only at three alternating locations of a hexagonal ring in the *h*-BN sheet. Essentially, a low % of carbon substitution into the *h*-BN sheet has little influence on the energy gap. On the other hand, several defect states appear above the top of VB in T-$B_2$ leading an energy gap of 0.446 eV, showing a big reduction in the energy gap relative to the pristine *h*-BN sheet. The reduction in the energy gap is associated with slight distortions of hexagonal rings at the boundary of the domain that manifest as defect states in the density of states (see, Fig. 3 (a)).

For hybrid sheets with Triangular-N graphene domains (red squares), the carbon atoms at the boundary of the domain are bonded with boron atoms only. Since boron atoms are electron deficient, they do not prefer to form $sp^2$ type of C-B bonds, but prefers to form a three-center two-electron type of B-B bonds with the next nearest neighbor boron atoms upon relaxation of the structure. Consequently, the symmetry of the graphene domain is broken and the domain boundary exhibits distortion, creating many defect states at the top of VB and at the bottom of the CB. The energy gap in such hybrid *h*-BN sheets show a drastic reduction in the energy gap compared to pristine *h*-BN sheets with gap values ranging from ~ 0.01-0.3 eV, except in the case of T-$N_1$ ($E_g$ ~ 0.9 eV) where the smallest triangular domain is not a carbon domain but a triangular domain containing boron atoms.

For circular graphene domains (green diamonds), the initial unrelaxed structures have the same number of C-B bonds and C-N bonds at the boundary. Since their bond lengths are different, one expects local strains at the domain boundary. However, these bonds are distributed uniformly and, therefore, the local strain arising from the differences in bond lengths is not that drastic upon relaxation. This may be the reason why the density of states is somewhat insensitive to the size of the circular graphene domain. The new states found in the gap must arise from states associated with C-B and C-N bonds along the boundary and the states associated with the graphene domains. For larger graphene domains, the contributions to these states come mainly from the carbon domain.

For the case of hexagonal hybrid *h*-BN/C sheets, the initial structures contain complete hexagonal carbon rings within the graphene domain. Since the six-fold symmetry of hexagonal rings is not broken upon relaxation, there are no symmetry-broken states within the gap. As the size of the carbon domain increases, the energy gap oscillates as the size of the graphene domain increases and this behavior is attributed to the aromaticity property of the carbon rings [17]. There are C-B bonds and C-N bonds along the domain boundary and the number of such bonds of each type are the same, and hence less distortions along the boundary. The contributions to the total density of states from C-B and C-N bonds are in general small compared to contributions from C-C bonds. The size of graphene domain dictates practically the energy gap.

For the case of rectangular graphene domains embedded in *h*-BN/C sheets, as the size of the domain increases, one observes the valence band edge to move towards the Fermi level and the conduction band edge moving away from the Fermi level. Additionally, one can see new shoulders near the conduction band edge. Since rectangular graphene domains are such that each of the opposite sides of the zigzag domain have different bonds, namely, C-N and C-B bonds, respectively, and, therefore, the perimeter of this domain is under local strain. However, on the armchair side of the domain, the C-N bond and C-B bonds alternate with each other. The net strain exhibited by the domain leads to the change of the energy gap of the hybrid *h*-BN/C sheet.

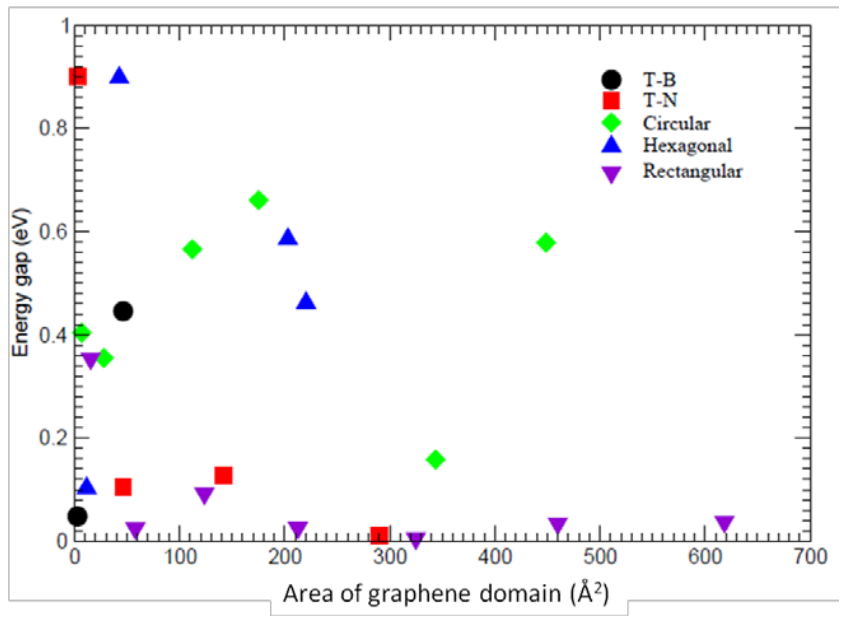

Figure 7 Energy gap versus area for graphene domains of different shapes and types (triangular, circular, hexagonal, or rectangular) embedded into *h*-BN sheets. An oscillatory behavior for $E_g$ versus area is obtained.

Table 2 Energy gaps of triangular boron (T-B), triangular nitrogen (T-N), circular (C), hexagonal (H) and rectangular (R) carbon domains in *h*-BN/C sheets.

| Graphene domains | Egap (eV) | Graphene domains | Egap (eV) | Graphene domains | Egap (eV) | Graphene domains | Egap (eV) |
|---|---|---|---|---|---|---|---|
| T-$B_1$ | 0.048 | C-1 | 0.404 | H-1 | 0.102 | R-1 | 0.354 |
| T-$B_2$ | 0.446 | C-2 | 0.355 | H-2 | 0.899 | R-2 | 0.025 |
| T-$N_1$ | 0.901 | C-3 | 0.566 | H-3 | 0.586 | R-3 | 0.092 |
| T-$N_2$ | 0.104 | C-4 | 0.661 | H-4 | 0.462 | R-4 | 0.026 |
| T-$N_3$ | 0.127 | C-5 | 0.158 | | | R-5 | 0.005 |
| T-$N_4$ | 0.009 | C-6 | 0.579 | | | R-6 | 0.034 |
| | | | | | | R-7 | 0.037 |

IV. Conclusion

In summary, by embedding carbon domains in *h*-BN sheets, their energy gaps can be manipulated. Both the carbon domain size and its shape influence the energy gap of pristine *h*-BN sheets. Since the properties contained in the database of the SCED-LCAO method mimics that of the DFT method, the absolute value of the energy gap obtained here must be used cautiously, but the relative trend obtained for the energy gap as a function of shape and size provides useful guidelines for materials design. Our results suggest that for controlled tuning of the energy gap of *h*-BN sheets, circular and hexagonal carbon domains are preferable.

Acknowledgements

We acknowledge computing resource support from the Cardinal Research Cluster at the University of Louisville. C. B. Kah acknowledges support received from the McSweeny Fellowship of the College of Arts and Sciences.